\documentstyle[12pt,epsf]{article}
\newcommand {\beq}{\begin{equation}}
\newcommand {\eeq}{\end{equation}}
\newcommand {\beqa}{\begin{eqnarray}}
\newcommand {\eeqa}{\end{eqnarray}}
\newcommand {\n}{\nonumber \\}

\begin{document}
\setlength{\oddsidemargin}{0cm}
\setlength{\baselineskip}{7mm}

\begin{titlepage}
 \renewcommand{\thefootnote}{\fnsymbol{footnote}}
$\mbox{ }$
\begin{flushright}
\begin{tabular}{l}
KEK-TH-683 \\
KUNS-1655 \\
April 2000
\end{tabular}
\end{flushright}

~~\\
~~\\
~~\\

\vspace*{0cm}
    \begin{Large}
       \vspace{2cm}
       \begin{center}
         {String Scale in Noncommutative Yang-Mills}
\\
       \end{center}
    \end{Large}

  \vspace{1cm}

\begin{center}

			Nobuyuki I{\sc shibashi}$^{1)}$\footnote
           {
e-mail address : ishibash@post.kek.jp},
           Satoshi I{\sc so}$^{1)}$\footnote
           {
e-mail address : satoshi.iso@kek.jp},
          Hikaru K{\sc awai}$^{2)}$\footnote
           {
e-mail address : hkawai@gauge.scphys.kyoto-u.ac.jp}{\sc and}
           Yoshihisa K{\sc itazawa}$^{1)}$\footnote
           {
e-mail address : kitazawa@post.kek.jp}

        $^{1)}$ {\it Laboratory for Particle and Nuclear Physics,\\
High Energy Accelerator Research Organization (KEK),}\\
               {\it Tsukuba, Ibaraki 305-0801, Japan} \\
        $^{2)}$ {\it Department of Physics, Kyoto University,
Kyoto 606-8502, Japan}\\
\end{center}

\vfill

\begin{abstract}
\noindent
We identify the effective string scale of noncommutative
Yang-Mills theory (NCYM) with the noncommutativity scale
through its dual supergravity description.
We argue that Newton's force law may be obtained with
4 dimensional NCYM with maximal SUSY.
It provides a nonperturbative compactification
mechanism of IIB matrix model.
We can associate NCYM
with the von Neumann lattice by the bi-local
representation.
We argue that it is superstring theory
on the von Neumann lattice.
We show that our identification of its effective string scale is
consistent with exact T-duality (Morita equivalence) of NCYM.

\end{abstract}

\vfill
\end{titlepage}
\vfil\eject

\section{Introduction}
\setcounter{equation}{0}

In recent studies
of nonperturbative aspects of superstring theory,
type IIB superstring is found to provide
the simplest setting\cite{IKKT}\cite{maldacena}.
However it is difficult
to obtain a realistic unified theory in IIB superstring
at least perturbatively.
Therefore we may expect that an entirely new type of
nonperturbative compactification
mechanism of IIB superstring exits\cite{AIKKT}\cite{review}.
On the other hand, a new compactification
mechanism which involves branes has been
proposed\cite{RS}.
Since branes naturally appear in superstring theory\cite{Polchinski},
such a mechanism is expected to apply for IIB superstring theory.

Noncommutative Yang-Mills theory (NCYM) has been obtained by compactifying
IIB matrix model on noncommutative tori\cite{CDS}.
We can simply obtain $\tilde{d}$ dimensional NCYM by expanding
IIB matrix model around $\tilde{d}$ dimensional noncommuting
backgrounds\cite{AIIKKT}.
In IIB matrix model, the dynamical variables
are the Hermitian matrices which are interpreted as the
space-time coordinates.
A $\tilde{d}$ dimensional noncommuting
background corresponds to a $\tilde{d}$ dimensional
noncommutative space-time.
The simplest idea for compactification in IIB matrix model is
to postulate that the compactification down to
$\tilde{d}$ dimensions is realized by expanding the model
around $\tilde{d}$ dimensional backgrounds.

In this paper we point out that
Newton's force law may be obtained with four dimensional NCYM
with maximal SUSY ($NCYM_4$).
Our argument is based on its dual supergravity
description\cite{hashimoto}\cite{ads+f}. We argue that there exists a massless
bound state in the effective Hamiltonian of supergravity which gives rise to
Newton's force law a la Randall and Sundrum.
Therefore $NCYM_4$ may be regarded as a four dimensional
compactification of IIB superstring.
The remarkable feature is that
it compactifies ten dimensional superstring
straight down to four dimensions.
The compactification of matrix models has been the outstanding
problem\cite{BFSS}\cite{WT}.
We argue that we can obtain four dimensional gauge theory and gravitation
with four dimensional noncommutative backgrounds in IIB matrix model.
In this sense, we have identified the most satisfactory
compactification mechanism of matrix models.
It is also possible to obtain $d+1$ dimensional NCYM theories
by expanding BFSS matrix model around $d$ dimensional noncommutative
backgrounds in an analogous way. It may be interesting to investigate
such theories through supergravity approach. However it is beyond
the scope of this paper.

In the large $N$ expansion of gauge theory, Feynman diagrams can be
classified with their world sheet topology.
This is a generic feature of matrix valued field theory.
On the other hand, string theory is perturbatively defined in terms of
field theory on the world sheet.
String theory may be nonperturbatively formulated
in the large $N$ limit of matrix models
in view of these remarkable correspondences.
IIB matrix model is such a proposal
which is a large $N$ reduced model of maximally supersymmetric
gauge theory\cite{IKKT}:
\beq
S  =  -{1\over g^2}Tr({1\over 4}[A_{\mu},A_{\nu}][A^{\mu},A^{\nu}]
+{1\over 2}\bar{\psi}\Gamma ^{\mu}[A_{\mu},\psi ]) .
\label{action}
\eeq
Here $\psi$ is a ten dimensional Majorana-Weyl spinor field, and
$A_{\mu}$ and $\psi$ are $N \times N$ Hermitian matrices.

With vanishing fermionic backgrounds,
the equations of motion are:
\beq
[A_{\mu},[A_{\mu},A_{\nu}]]=0.
\label{SEOM}
\eeq
The following solutions correspond to BPS-saturated backgrounds:
\beq
[A_\mu,A_\nu] =c-number\equiv C_{\mu\nu} .
\eeq
Since we interpret $A_{\mu}$ as space-time coordinates
due to $\cal{N}$=2 SUSY,
we expect to obtain $\tilde{d}$ dimensional space-time
with $\tilde{d}$ dimensional solutions of this type.
We further expect to obtain $\tilde{d}$ dimensional gauge theory.
Since matrices form noncommutative but associative algebra,
we expect a deep connection to noncommutative geometry
\cite{Connes}.
In fact we have obtained NCYM of 16 supercharges with these
backgrounds\cite{AIIKKT}.
Ordinary gauge theory appears as the low energy effective theory.
Since short open strings correspond to gauge particles, we
indeed find another evidence that IIB matrix model can describe
infinite numbers of fundamental strings.
We have further pointed out that NCYM contains
nonlocal degrees of freedom which may be interpreted as long
open strings\cite{IIKK}\cite{IKK}. We have indeed shown that
they give rise to gravitational interactions at the one loop level
as it is expected in superstring theory\cite{AIIKKT}\cite{IKK}.

Since NCYM seems to contain the both gauge theory and gravitation,
it is very likely that it is equivalent to
superstring theory in a particular background.
The major issue here is the renormalizability of NCYM
\cite{ChRo}\cite{MRS}.
We have shown that the high energy behavior of NCYM is
equivalent to large $N$ gauge theory by using the bi-local
field representation\cite{IKK}.
Although it also exhibits long range interactions
which we interpret as gravitation,
it is very likely that NCYM exits at least for
$\tilde{d} \leq 4$.

NCYM is often argued to be the low energy limit of
string theory with constant $b_{\mu\nu}$ field
\cite{Cheung:1998nr}\cite{Ardalan:1999ce}\cite{Chu:1999qz}
\cite{Schomerus:1999ug}\cite{SW} .
However long range interactions
are found due to the presence of
long `open strings' which might signal the presence
of `closed strings'
\cite{AIIKKT}\cite{IKK}\cite{MRS}\cite{hayakawa}\cite{Fischler}\cite{RS2}.
These issues are currently under active investigations
\cite{Andreev}\cite{Kiem}\cite{Bilal}\cite{Gomis}\cite{Liu}.
In this paper we propose that NCYM is
superstring theory on the von Neumann lattice whose effective
string scale $\alpha '_{eff}$ is set by $C_{\mu\nu}$.

The organization of this paper is as follows.
In section 2, we argue that Newton's force law
is obtained with $NCYM_4$.
Since it is a nonperturbative problem, we study its
dual supergravity description.
In section 3, we briefly summarize our formulation
of NCYM as twisted reduced models.
In section 4, we estimate the string tension
of NCYM using the formalism of section 3.
We find that our estimate is consistent with
string theoretic expectations in section 2.
In section 5, we investigate the graviton exchange
process by the one loop perturbation theory.
We conclude in section 6 with discussions.

\section{$NCYM_4$ as a unified theory}
\setcounter{equation}{0}

In this section, we argue that $NCYM_4$
contains four dimensional gauge theory and gravitation.
It is clear that NCYM contains ordinary gauge theory
since the noncommutative
phases become ineffective at tree level
in the low energy limit.
The remarkable possibility is that it may also contain gravitation.
We first observe the long range interaction
at the one loop level which is specific in NCYM.
It can be interpreted as gravitational interaction
in IIB superstring as it is explained in section 5.
In string theory, closed string exchanges should be visible at open
string one loop level.
Therefore this phenomenon is another stringy feature
of NCYM.
Although we can see the glimpse of closed strings at the
one loop level, we need to understand the quantum effects
to all orders to investigate the gravitational sector
of $NCYM_4$.

In order to study such a problem,
we recall the supergravity solution
of $m$ coincident D3-branes with the constant NS B field strength
$b$\cite{ads+f}:
\beqa
e^{\phi}&=&g_{\infty}{(1+{g_{\infty}m\alpha'^2(1+\alpha'^2b^2)\over r^4})
\over (1+{g_{\infty}m\alpha'^2\over r^4})} ,\n
{1\over \alpha'}ds^2&=&{1\over\alpha'}
(1+{g_{\infty}m\alpha'^2(1+\alpha'^2b^2)\over r^4})^{1\over 2}
({d\vec{x}^2\over 1+{g_{\infty}m\alpha'^2\over r^4}}
+ dr^2 +r^2d\Omega_5^2 ) ,\n
B_2&=&{b\over (1+{g_{\infty}m\alpha'^2\over r^4})}dx\wedge dy
+{b\over (1+{g_{\infty}m\alpha'^2\over r^4})}dz\wedge d\tau ,\n
C_2&=&i{1\over g_{\infty}(1+\alpha'^2b^2)}B_2 ,\n
C_0&=& -i{b^2\over g_{\infty}(1+\alpha'^2b^2)}
{1\over (1+{g_{\infty}m(1+\alpha'^2b^2)\over r^4})} ,\n
F_{0123r}&=&-4i{1\over (1+{g_{\infty}m\alpha'^2\over r^4})^2}
{{m\over r^5}} .
\label{ads}
\eeqa
Here $g_{\infty}$ is the dilaton expectation value at $r={\infty}$.
$\vec{x}$ denotes four dimensional space-time coordinates in this
section.

These background fields appear in
the Euclidean IIB supergravity action:
\beqa
S_{IIB}&=&S_{NS}+S_{R}+S_{CS},\n
S_{NS}&=&-{1\over 2}\int d^{10}x\sqrt{g}
e^{-2\phi}(R+4\partial_{\mu}\phi\partial^{\mu}\phi
-{1\over 2}{H_3}^2),\n
S_{R}&=&{1\over 4}\int d^{10}x\sqrt{g}
({F_1}^2+{\tilde{F}_3 }^2+{1\over 2}{\tilde{F}_5 }^2),\n
S_{CS}&=&{1\over 4}\int C_4\wedge H_3\wedge F_3 ,
\eeqa
where
\beqa
\tilde{F}_3&=&F_3-C_0\wedge H_3,\n
\tilde{F}_5&=&F_5-{1\over 2}C_2\wedge H_3+{1\over 2}B_2\wedge F_3 .
\eeqa

We identify the $r$ dependent metric $g_{\alpha\beta}$ in
eq.(\ref{ads}) as the four dimensional metric for fundamental strings:
\beq
g_{\alpha\beta}
=(1+{g_{\infty}m\alpha'^2(1+\alpha'^2b^2)\over r^4})^{1\over 2}
({1\over 1+{g_{\infty}m\alpha'^2\over r^4}})\delta_{\alpha\beta} .
\eeq
We postulate that D3-branes are located at the maximum
of $g_{\alpha\beta}$ , namely at the `boundary'
$r=(g_{\infty}m\alpha'^2)^{1/4}$. Since open strings live on the
D3-branes, we identify the open string metric $G_{\alpha\beta}$ with
$g_{\alpha\beta}$ at the `boundary' as
\beqa
G_{\alpha\beta}
\sim (1+\alpha'^2b^2)^{1\over 2}\delta_{\alpha\beta} .
\label{metricrel2}
\eeqa

Eq.(\ref{ads}) indicates that fundamental string metric
grows at smaller $r$.
This phenomenon may be interpreted that closed strings become
dynamical due to the quantum effects in NCYM.
The graviton exchanges we find at the one loop level
in section 5 support such an interpretation.
We consider the case that the noncommutativity scale $l_{NC}$ is much smaller
than the string scale $l_s$. Let us focus on the physics at the
noncommutativity
scale by letting $\alpha' \rightarrow \infty$ but keeping $b,r\sim O(1)$.
Although it might appear to be a strange limit,
it is equivalent to consider the standard $\alpha ' \sim \epsilon^{1/2},
x_{\mu} \sim \epsilon^{1/2} \tilde{x}_{\mu}, b \sim \epsilon^{-1}$ limit.
The remarkable point is that open string metric is
given by eq.(\ref{metricrel2}) as $\alpha' b$ in this limit.

The Polyakov action for fundamental strings becomes in such a limit as
\beqa
&&{1\over \alpha '}\int d^2zG_{\alpha\beta}\partial x^{\alpha}
\bar{\partial} x^{\beta}
+\int d^2z  b_{\alpha\beta}\partial_0 x^{\alpha}\partial_1 x^{\beta}
+\cdots\n
&\sim&
{b}\int d^2z (\partial_0 x_{\alpha}\partial_0 x^{\alpha}
-\partial_1 x_{\alpha}\partial_1 x^{\alpha})
+\int d^2z  b_{\alpha\beta}\partial_0 x^{\alpha}\partial_1 x^{\beta}
+\cdots .
\eeqa
So the Hamiltonian for open strings behaves like
\beq
cp^2+ \sum_{k\neq 0} kn_k ,
\label{openprop}
\eeq
where
$c=1/b$ and $n_k$ denote the number operators of the oscillator modes.
Here we find that the noncommutativity scale now acts as the effective
string scale!

Supergravity description of $U(m)$ $NCYM_4$ may be obtained
by considering large $\alpha'$ and small $g_{\infty}$ limit while keeping
$\lambda=g_{\infty}\alpha'^2b^2$ fixed\cite{ads+f}:

\beqa
e^{\phi}&=&
({ m\lambda^2\over r^4})
{1\over (1+{m\lambda  \over r^4})} ,\n
{1\over \alpha'}ds^2&=&
({m\lambda\over r^4})^{1\over 2}(
{d\vec{x}^2\over
1+{m\lambda  \over r^4}}+
dr^2 +r^2d\Omega_5^2 ) ,\n
B_2&= &{1\over (1+{m\lambda \over r^4})}dx\wedge dy +
{1\over (1+{m\lambda \over r^4})}dz\wedge d\tau ,\n
C_2&=&i{1\over \lambda} B_2 ,\n
C_0&=& -i{r^4\over m\lambda^2} ,\n
F_{0123r}&=&
-4i{1\over (1+{m\lambda \over r^4})^2}{m\over r^5}
\label{adsrs} .
\eeqa
Here we have also put $b=1$ which implies that
the noncommutativity scale $l_{NC}$ is $O(1)$.
Since we are looking at the vicinity of
the D3-branes in this limit, we expect to find massless open strings.
However we also find oscillator modes since the effective string scale is
set by $l_{NC}$ as in eq.(\ref{openprop}).

We recall that there is a crossover at the noncommutativity scale $l_{NC}$
in NCYM.
When we consider the Wilson loops, we find that
the planar diagrams dominate at larger momentum scale than $1/l_{NC}$ and the
diagrams of all topology contribute in the opposite limit\cite{twisted}.
It may be interpreted that the string coupling (dilaton expectation value)
is scale dependent.
It is because in our IIB matrix model conjecture,
the tree level string theory is considered to be obtained by
summing planar diagrams and string perturbation theory
is identified with the topological expansion of the matrix model.
With this interpretation, the string coupling
grows as the relevant momentum scale is decreased while it vanishes
in the opposite limit. In the $D$ brane interpretation,
the small momentum region
corresponds to the vicinity of the brane, while the large momentum
region corresponds to
the region far from the brane since the Higgs expectation value
plays the same role with the
momentum scale.
In this sense $e^{\phi}$ in eq.(\ref{adsrs}) behaves just like the $NCYM_4$
\cite{IIKK}\cite{MRS}.

The small $r$ behavior of eq.(\ref{adsrs}) is
identical with ordinary $AdS$/CFT correspondence
if we identify $\lambda$ as the coupling of ordinary
Yang-Mills theory\cite{maldacena}.
This result is reasonable since the low energy limit
of $NCYM_4$ contains ordinary gauge theory with precisely the same
relation between the coupling constants.
It is because in string theory the coupling of $NCYM_4$
is given by $\lambda=g_\infty \alpha'b^2$ when $\alpha'$ is large.

We may now resort to the standard argument to
justify the supergravity description as follows.
Since $m\lambda$ sets the radius of `$AdS_5$' and $S_5$,
supergravity description is valid
in the strong 't Hooft coupling limit of $NCYM_4$.
The mass scale for the Kaluza-Klein modes can be estimated to be of order
$1/(m\lambda)^{1/4}l_{NC}$.
We need to consider large $m$ limit
also in order to keep the dilaton expectation value to be small.
As we have argued, the mass scale of the oscillator modes is
set by the effective string scale as $1/l_{NC}$.

In order to investigate the gravitational interaction,
we introduce external energy momentum tensor $T_{\mu\nu}$.
As an explicit example, we may consider the photon-photon
scattering on the `brane' as in section 5.
Having such a case in mind, we assume that the indices
of the nonvanishing components of $T_{\alpha\beta}$ run over four
dimensional space-time coordinates. We also assume that
it is traceless in the four dimensional subspace.
There is an ambiguity concerning its dilaton dependence.
It may be natural to assume that it contains the factor
$e^{-\phi}$ from string theory point of view.
However such an ambiguity does not change the main conclusions
in this section.

We may adopt the coordinate system where
the five dimensional subspace $(\vec{x},\rho)$ is conformally flat
\beq
{1\over \alpha'}ds^2=(m\lambda)^{1\over 2}(A(\rho )(d \vec{x}^2 +d
\rho^2)+d\Omega_5^2) .
\label{confmet}
\eeq
Since
\beq
\rho =\int dr\sqrt{1+{m\lambda\over r^4}},
\eeq
we find that
\beq
A(\rho )\sim 1/\rho ^2 ,~~\rho \rightarrow \pm \infty .
\eeq
It has the unique maximum at $\rho=0$  ($r=(m\lambda)^{1/4}$).
It is illustrated in Figure 1.
\begin{figure}[h]
\begin{center}
\leavevmode
\epsfxsize=6cm
\epsfbox{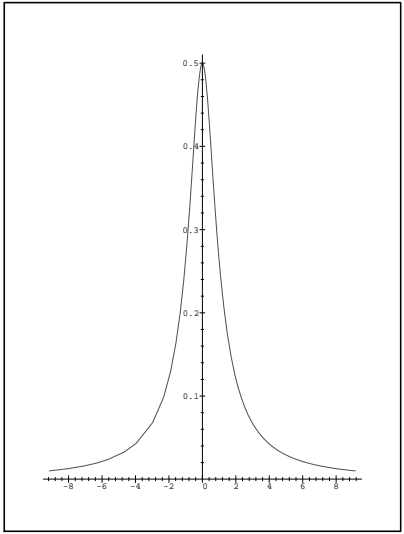}
\caption{Conformal factor $A(\rho)$ is plotted as a function of $\rho$.
$m \lambda$ is set to 1.
}
\label{fig:conformal}
\end{center}
\end{figure}
Our strategy is to expand the metric and equations of motion
around the classical solution to the first order of the fluctuation.
In the following investigation,
we use the formalism developed in \cite{KKN}.
\footnote{We use the sign conventions for the curvature
tensors of Misner, Thorn and Wheeler in this paper
while those of 't Hooft and Veltman\cite{tHooft} have been used in \cite{KKN}.
Although the signs of the Riemann tensor are the same,
those of the Ricci tensor and scalar curvature are the opposite
in these conventions.}
We parametrize the metric as $g_{\mu\rho}{(e^h)^{\rho}}_{\nu}$ where
$g_{\mu\nu}$ is the background metric and ${h^{\rho}}_{\rho}=0$
(traceless). The tensor indices are raised and lowered by
the background metric. This formalism explicitly separates
${h^{\mu}}_{\nu}$ from the conformal mode of the metric.

The equation of the motion with respect to ${h^{\mu}}_{\nu}$ is
\beq
 {R^{\mu}}_{\nu}+2\nabla^{\mu}\nabla_{\nu}\phi
-{1\over 2}{H_3^{\mu}}{H_{3\nu}} +\cdots
=\kappa^2{T^{\mu}}_{\nu} ,
\label{Einstein}
\eeq
where $\kappa =e^{\phi}$. We have suppressed the
contributions from the R-R sector.
The advantage to consider the four dimensional traceless
energy momentum tensor ${T^{\mu}}_{\nu}$ is that all other equations of
motion are satisfied to the first order of ${h^{\mu}}_{\nu}$.
In this sense it minimally excites gravitons.

Eq. (\ref{Einstein}) is expanded
to the first order of ${h^{\mu}}_{\nu}$ as
\beqa
&&{1\over 2}
\nabla^{\rho}\nabla_{\rho}{h^{\mu}}_{\nu}
-{1\over 2}h^{\mu\rho},_{\rho\nu}
-{1\over 2}h_{\nu\rho},^{\rho\mu}
\n
&&-{h^{\mu}}_{\nu},^{\rho}\partial_{\rho}\phi
-{1\over 2}{H^{\mu}}_{\rho\tau}
{H_{\nu}}^{\sigma\tau}{h^{\rho}}_{\sigma}
-{1\over 4}{H^{\rho}}_{\sigma\tau}
{H_{\nu}}^{\sigma\tau}{h^{\mu}}_{\rho}
+\cdots \n
&=& -\kappa^2{T^{\mu}}_{\nu} .
\label{fulleqn}
\eeqa
In the coordinate system of eq.(\ref{confmet}),
it is consistent to assume that the tensor
indices of the nonvanishing ${h^{\alpha}}_{\beta}$
resides in the four dimensional space-time
since the tensor indices of ${T^{\alpha}}_{\beta}$ are also four dimensional.
It is also consistent to assume that ${h^{\alpha}}_{\alpha}=0$ since
${T^{\alpha}}_{\alpha}=0$.
We also adopt the $h^{\alpha\gamma},_{\gamma}=0$ gauge.


Our strategy is to first study
the following free equation of motion for ${{h}^{\alpha}}_{\beta}$
in the ten dimensional curved  space-time:
\beq
{1\over 2}
\nabla^{\mu}\nabla_{\mu}{{h}^{\alpha}}_{\beta}
=-\kappa^2 {T^{\alpha}}_{\beta} .
\label{freeeq}
\eeq
Eq.(\ref{freeeq}) can be rewritten as:
\beqa
{H}{{h}^{\alpha}}_{\beta}&=&2A\kappa^2 {T^{\alpha}}_{\beta},\n
{H}&=&-{A\over \sqrt{g}}\partial_{\mu}\sqrt{g}g^{\mu\nu}\partial_{\nu} .
\label{Shrod1}
\eeqa
The Hamiltonian is
\beq
H=-{1\over (m\lambda)^{1\over 2}}(\vec{\nabla}^2+{\partial ^2 \over \partial
\rho^2} +{3\over 2}{A'\over A}{\partial\over \partial \rho}+A\hat{L}^2) ,
\label{Hamilton1}
\eeq
where $A'={\partial A/\partial \rho}$.
The symbols $\vec{\nabla}^2$ and $\hat{L}^2$ denote
the Laplacians on $R^4$ and $S^5$ respectively.
We can further simplify eq. (\ref{Shrod1})
by the similarity transformation as
\beqa
\tilde{H}A^{{3\over 4}}h_{\alpha\beta}&=
&2\kappa^2 A^{7\over 4}T_{\alpha\beta} ,\n
\tilde{H}&=& A^{3/4} H A^{-3/4}\n
&=& -{1\over (m\lambda)^{1\over 2}}
(\vec{\nabla}^2+{\partial ^2 \over
\partial \rho^2} -{3\over 4}{A''\over A}+{3\over 16}{A'^2\over A^2}
+A\hat{L}^2) .
\eeqa

We concentrate on the $S$ wave on $S^5$
in what follows.
The eigenfunction of $\tilde{H}$ is found to be
$exp(i\vec{k}\cdot\vec{x})\tilde{\phi}_l$ with the eigenvalue
${(m\lambda)^{-1\over 2}}(\vec{k}^2+E_l)$.
Note that $E_l$ acts as the four dimensional mass of the
various modes.
It can be obtained by solving the following quantum mechanics problem
\beq
(-{\partial ^2 \over \partial \rho^2}
+{3\over 4}{A''\over A}-{3\over 16}{A'^2\over A^2}) \tilde{\phi}
=E \tilde{\phi} .
\label{Hamilton2}
\eeq
Let us introduce a super-charge
\beq
Q={\partial \over i\partial\rho}\sigma_1+{3\over 4}{A'\over A}\sigma_2 .
\eeq
Since the relevant Hamiltonian can be embedded in $Q^2$, we only need to solve
$Q \tilde{\phi}=0$ to find the zeromodes.
The solution is
\beq
\tilde{\phi} = e^{{3\over 4} \int d\rho ({A'\over A})} .
\eeq
We conclude that there is a single zero energy bound state
with the conformal factor $A$ of our type as follows
\beq
\tilde{\phi}_0 = A^{3\over 4} .
\eeq
Such a zero mode corresponds to a massless field
in four dimensions.

The propagator in this basis is
\beq
G(x,y)=\sum_n <x|n>{1\over E_n}<n|y> ,
\eeq
where $|n>$ is the eigenstate of $\tilde{H}$ with the
eigenvalue $E_n$.
We may adopt the vacuum saturation type approximation
by only considering $<x|n>=exp(i\vec{k}\cdot\vec{x}) \tilde{\phi}_0$
as the intermediate states.
In this way we obtain the propagator of massless fields in
four dimensions:
\beqa
G(x,y)&\sim&\int d^4k exp(i\vec{k}\cdot(\vec{x}-\vec{y}))
{1\over {k}^2}\tilde{\phi}_0 (\rho)\tilde{\phi}_0(\rho')\n
&\sim&{1\over (\vec{x}-\vec{y})^2}\tilde{\phi}_0 (\rho)
\tilde{\phi}_0 (\rho') .
\eeqa
As the final result of these investigations,
we find the following gravitational interaction:
\beqa
&&-{1\over 2}\int d^{10}xA^{5\over 2}(\rho)
{T_{\mu}}^{\nu}(\vec{x},\rho){h^{\mu}}_{\nu}(\vec{x},\rho)\n
&=&-\int d^{10}x\int d^{10}yA^{7\over 4}(\rho)
{T_{\alpha}}^{\beta}(\vec{x},\rho)\tilde{\phi}_0 (\rho)
{1\over (\vec{x}-\vec{y})^2}\tilde{\phi}_0 (\rho')
\kappa^2(\rho'){T^{\alpha}}_{\beta}(\vec{y},\rho')
A^{7\over4}(\rho') \n
&\sim&
-\bar{\kappa}^2\int d^{4}x\int d^{4}y
{\tilde{T}_{\alpha}}^{\beta}(\vec{x}){\tilde{T}^{\alpha}}_{\beta}(\vec{y})
{1\over (\vec{x}-\vec{y})^2} .
\label{newton}
\eeqa
We have introduced the four dimensional energy momentum tensor
\beq
{\tilde{T}^{\alpha}}_{\beta}(\vec{x})
=\int d\rho d\Omega_5 {T^{\alpha}}_{\beta}(\vec{x},\rho)A^{5\over 2}(\rho) .
\eeq
The interaction between them is of the four dimensional
graviton exchange type with the gravitational coupling
$\tilde{\kappa}=\kappa(0)$.

Here we remark on the Hermiticity of the Hamiltonians
in eqs. (\ref{Hamilton1}) and (\ref{Hamilton2}).
The latter is  Hermitian with respect to the trivial
norm since
\beqa
&&\int d\rho \tilde{\phi}
({\partial \over i\partial \rho}-i{3\over 4}{A'\over A})
({\partial \over i\partial \rho}+i{3\over 4}{A'\over A})\tilde{\phi} \n
&=&
\int d\rho
(-{\partial \over i\partial \rho}-i{3\over 4}{A'\over A})\tilde{\phi} \
({\partial \over i\partial \rho}+i{3\over 4}{A'\over A}) \tilde{\phi} .
\eeqa
It is translated into the Hermiticity condition on the former
after the similarity transformation as
\beq
\int d\rho A^{3\over 2}
(-{\partial \over i\partial \rho})\phi
({\partial \over i\partial \rho}) \phi\n
\sim  \int d\rho \sqrt{g} g^{\rho\rho}
{\partial \over \partial \rho}\phi
{\partial \over \partial \rho} \phi ,
\eeq
where $\phi = A^{-3/4} \tilde{\phi}$.
Therefore our Hamiltonian is positive definite with respect
to the natural norm defined by the string frame metric $g_{\mu\nu}$.
In this sense our important physical input is our
identification of the physical metric with string frame
metric $g_{\mu\nu}$.
We have checked that our zero mode remains the exact zero mode
after taking account of other terms in eq.(\ref{fulleqn}).

Therefore we argue that we can obtain four dimensional
gravity with $NCYM_4$
a la Randall-Sundrum\cite{RS}.
Since not only the metric but also the dilaton expectation value
(string coupling)
rapidly decay in the large $r$ region, we expect that there is
essentially nothing outside the noncommutativity scale
transverse to the `brane'.
In fact we have postulated this kind of `compactification'
mechanism in the matrix models
\cite{AIKKT}\cite{review}\cite{Nishimura}.
We have expected that four dimensional gravitation
is obtained if the eigenvalue distribution of the matrices
are four dimensional. It is because
the matrices represent space-time coordinates in our proposal.
In our interpretation, there is simply nothing
outside the support of the eigenvalue distributions,
not even space-time.

We observe that
this Euclidean solution can be analytically continued
into Minkowski space-time only in the small $r$ region.
One possible interpretation of such a solution is to maintain that
Minkowski space-time appears from $NCYM_4$ as its low energy
approximation.  We may identify the
noncommutativity scale with Planck scale if we apply this model to our
space-time. Although the Lorentz invariance is broken at the
noncommutativity scale in this model, such a possibility is not excluded
by the experiments. Therefore $NCYM_4$ is a candidate of the unified
theory of interactions. We explain in the subsequent sections that IIB
matrix model naturally provides us with such a theory.
We still need to solve many problems such as
breaking SUSY and finding chiral fermions
to construct a realistic unified theory.
We hope that these problems can be solved by further
investigations in IIB matrix model.

\section{Noncommutative field theories as twisted reduced models}
\setcounter{equation}{0}

In this section we briefly recapitulate
our formulation of NCYM through large $N$ reduced models.
We have pointed out that well-known twisted reduced
models\cite{RM}\cite{twisted}
\footnote{The relevance of reduced models and string theory
was first recognized in\cite{Zacos}\cite{Bars}.}
are equivalent to NCYM.
This connection is further studied in\cite{AMNS}\cite{bars99}\cite{AMNS1}.
We consider $d$ dimensional $U(n)$ gauge theory coupled to
adjoint matter as an example:
\beq
S=-\int d^dx {1\over g^2}Tr({1\over 4}[D_{\mu},D_{\nu}][D_{\mu},D_{\nu}]
+{1\over 2}\bar{\psi}\Gamma _{\mu}[D_{\mu},\psi ]) ,
\eeq
where $\psi$ is a Majorana spinor field.
The corresponding reduced model is
\beq
S=- {1\over g^2}Tr({1\over4}[A_{\mu},A_{\nu}][A_{\mu},A_{\nu}]
+{1\over 2}\bar{\psi}\Gamma _{\mu}[A_{\mu},\psi ]) .
\eeq
Now $A_\mu$ and $\psi$ are $n\times n$ Hermitian matrices
and each component of $\psi$ is $d$-dimensional Majorana-spinor.

We expand $A_{\mu}=\hat{p}_{\mu}+\hat{a}_{\mu}$
around the following classical solution
\beq
[\hat{p}_{\mu},\hat{p}_{\nu}]=iB_{\mu\nu} ,
\eeq
where  $B_{\mu\nu}$ are $c$-numbers.
We assume the rank of $B_{\mu\nu}$
to be $\tilde{d}$ and define its inverse $C^{\mu\nu}$ in $\tilde{d}$
dimensional subspace.
The directions orthogonal
 to the subspace
is called the transverse directions.
$\hat{p}_{\mu}$ satisfy the canonical commutation relations and
they span the $\tilde{d}$ dimensional phase space.
The semiclassical correspondence shows that the
volume of the phase space is $V_p=n(2\pi)^{\tilde{d}/2} \sqrt{detB}$.

We Fourier decompose
$\hat{a}_{\mu}$ and $\hat{\psi}$ fields as
\beqa
\hat{a}&=&\sum_k \tilde{a}(k) exp(iC^{\mu\nu}k_{\mu}\hat{p}_{\nu}) ,\n
\hat{\psi}&=&\sum_k \tilde{\psi}(k) exp(iC^{\mu\nu}k_{\mu}\hat{p}_{\nu}) ,
\label{twist}
\eeqa
where $exp(iC^{\mu\nu}k_{\mu}\hat{p}_{\nu})$ is the eigenstate of
adjoint $P_{\mu}=[\hat{p}_{\mu},~]$ with the eigenvalue $k_{\mu}$.
The Hermiticity requires that $\tilde{a}^* (k)=\tilde{a}(-k)$ and
$\tilde{\psi} ^*(k)=\tilde{\psi} (-k)$.

We can construct a map from a matrix to a function
as
\beq
\hat{a} \rightarrow a(x)=\sum_k \tilde{a}(k) exp(ik\cdot x) ,
\label{proj}
\eeq
where $k\cdot x=k_{\mu}x^{\mu}$.
By this construction, we obtain the $\star$ product
\beqa
\hat{a}\hat{b} &\rightarrow& a(x)\star b(x),\n
a(x)\star b(x)&\equiv&exp({C^{\mu\nu}\over 2i}{\partial ^2\over
\partial\xi^{\mu}
\partial\eta^{\nu}})
a(x+\xi )b(x+\eta )|_{\xi=\eta=0} .
\label{star}
\eeqa
The operation $Tr$ over matrices can be exactly mapped onto the integration
over functions as
\beq
Tr[\hat{a}] =
\sqrt{det B}({1\over 2\pi})^{\tilde{d}\over 2}\int d^{\tilde{d}}x a(x) .
\label{traceint}
\eeq
The twisted reduced model can be shown to be equivalent to
NCYM by the
the following map from matrices onto functions
\beqa
\hat{a} &\rightarrow& a(x) , \n
\hat{a}\hat{b}&\rightarrow& a(x)\star b(x) ,\n
Tr&\rightarrow&
\sqrt{det B}({1\over 2\pi})^{\tilde{d}\over 2}\int d^{\tilde{d}}x .
\label{momrule}
\eeqa
The following commutator is mapped to the covariant derivative:
\beq
[\hat{p}_{\mu}+\hat{a}_{\mu},\hat{o}]\rightarrow
{1\over i}\partial_{\mu}o(x)+a_{\mu}(x)\star o(x)-o(x)\star a_{\mu}(x)
\equiv [D_{\mu},o(x)] ,
\label{pcovder}
\eeq
We may interpret  the newly emerged coordinate space
as the semiclassical limit of $\hat{x}^{\mu}=C^{\mu\nu}\hat{p}_{\nu}$.
Therefore we can interpret $A_{\mu}$ as momenta as well
in IIB matrix model with noncommutative backgrounds since
$\hat{x}$ and $\hat{p}$ are linearly related. It is the
reflection of the remarkable T-duality property of the theory.
The space-time translation is realized by the following unitary
operator:
\beq
exp(i\hat{p}\cdot d)\hat{x}^{\mu}
exp(-i\hat{p}\cdot d)\n
=\hat{x}^{\mu}+d^{\mu} .
\label{transl}
\eeq

Applying the rule eq.(\ref{momrule}), the bosonic action becomes
\beqa
&&-{1\over 4g^2}Tr[A_{\mu},A_{\nu}][A_{\mu},A_{\nu}]\n
&=&
{\tilde{d}nB^2\over 4g^2}-\sqrt{det B}({1\over 2\pi})^{\tilde{d}\over 2}
\int d^{\tilde{d}}x
{1\over g^2} ({1\over 4}
[D_{\alpha},D_{\beta}][D_{\alpha},D_{\beta}]\n
&&+{1\over 2}[D_{\alpha},
\varphi_{\nu}][D_{\alpha},\varphi_{\nu}]
+{1\over 4}[\varphi_{\nu},\varphi_{\rho}]
[\varphi_{\nu},\varphi_{\rho}])_{\star} .
\eeqa
In this expression, the indices $\alpha,\beta$ run over $\tilde{d}$
dimensional world volume directions  and $\nu,\rho$
over the transverse directions.
We have replaced $a_{\nu}\rightarrow\varphi_{\nu}$ in the transverse
directions. Inside $(~~)_{\star}$, the products should be understood as
$\star$
products and hence commutators do not vanish.
The fermionic action becomes
\beqa
&&{1\over g^2}Tr\bar{\psi}{\Gamma}_{\mu}[A_{\mu},\psi]\n
&=&
\sqrt{det B}({1\over 2\pi})^{\tilde{d}\over 2}
\int d^{\tilde{d}}x{1\over g^2}
(\bar{\psi}{\Gamma}_{\alpha}[D_{\alpha},\psi ]
+\bar{\psi}\Gamma_{\nu}[\varphi_{\nu},\psi ])_{\star} .
\eeqa
We therefore find noncommutative U(1) gauge theory.
In order to obtain NCYM with $U(m)$ gauge group,
we need to consider new classical solutions which are obtained by replacing
each element
of $\hat{p}_{\mu}$ by the $m\times m$ unit matrix:
\beq
\hat{p}_{\mu} \rightarrow \hat{p}_{\mu}\otimes {\mathbf{1}}_m .
\eeq

The Hermitian models are invariant under the unitary transformation:
$A_{\mu}\rightarrow UA_{\mu}U^{\dagger},
\psi \rightarrow U\psi U^{\dagger}$. As we shall see, the gauge
symmetry can be embedded in the $U(N)$ symmetry.
We expand $U=exp(i\hat{\lambda} )$ and parameterize
\beq
\hat{\lambda}=\sum_k \tilde{\lambda} (k)
exp(i{k}\cdot\hat{x}) .
\eeq
Under the infinitesimal gauge transformation, we find the fluctuations around
the fixed background transform as
\beqa
\hat{a}_{\mu}&\rightarrow & \hat{a}_{\mu}+i[\hat{p}_{\mu},\hat{\lambda} ]
-i[\hat{a}_{\mu},\hat{\lambda} ] ,\n
\hat{\psi}&\rightarrow & \hat{\psi} -i[\hat{\psi},\hat{\lambda} ] .
\eeqa
We can map these transformations
 onto the gauge transformation
in NCYM by our rule eq.(\ref{momrule}):
\beqa
&&a_{\alpha}(x) \rightarrow a_{\alpha}(x) +
{\partial \over \partial x^{\alpha}}\lambda (x)
-ia_{\alpha}(x)\star \lambda (x)+i\lambda (x)\star a_{\alpha}(x) ,\n
&&\varphi_{\nu}(x) \rightarrow \varphi_{\nu}(x)
-i\varphi_{\nu}(x)\star \lambda (x)+i\lambda (x)\star \varphi_{\nu}(x) ,\n
&&\psi (x)\rightarrow \psi (x)
-i\psi (x)\star \lambda (x)+i\lambda (x)\star \psi (x) .
\eeqa

We have introduced another representation of matrices\cite{IKK}.
For simplicity we consider the two dimensional case first:
\beq
[\hat{x}, \hat{y}] = -iC .
\eeq
This  commutation relation is realized by the guiding center coordinates
of  the two dimensional system of electrons
in magnetic field.
We recall that we have $n$ quanta  with $n$ dimensional matrices.
Each quantum occupies the space-time volume of $2\pi C$.
We may consider a square von Neumann lattice with the lattice
spacing $l_{NC}$ where $l_{NC}^2=2\pi C$.
This spacing $l_{NC}$ gives the noncommutative scale.
Let us denote the most localized state centered at the origin
by $|0 \rangle$.
It is annihilated by the operator $\hat{x}^{-}=\hat{x}-i\hat{y}$.
We construct states localized around each lattice site
by utilizing translation operators
$|x_{i} \rangle =exp(-ix_i \cdot\hat{p})|0 \rangle$.
They are the coherent states on a von Neumann lattice
${\bf x}_i = l_{NC} (n_i {\bf e}^x  + m_i {\bf e}^y)$
where $n,m \in {\bf Z}.$
The generalizations to arbitrary even $\tilde{d}$ dimensions are
straightforward.
\par
We evaluate the following matrix elements
\beq
\rho_{ij} \equiv \langle x_i|x_j \rangle =exp({i\over 2}B_{\mu\nu}
x_i^{\mu}x_j^{\nu})exp
(-{(x_i-x_j)^2\over 4 C}).
\eeq
Although  $|x_i \rangle $ are non-orthogonal,
$\langle x_i|x_j \rangle $ exponentially vanishes
when $(x_i-x_j)^2$ gets large.
We also find
\beq
\langle x_i|exp(ik\cdot\hat{x})|x_j \rangle =
exp(ik\cdot{(x_i+x_j)\over 2}+{i\over 2}B_{\mu\nu}x_i^{\mu}x_j^{\nu})
exp(-{(x_i-x_j-d)^2\over 4 C}) ,
\eeq
where $d^{\mu}=C^{\mu\nu}k_{\nu}$.
This matrix element sharply peaks at $x_i-x_j=d$. It supports our
interpretation that
the eigenstate $exp(ik\cdot\hat{x})$ with $|k_{\mu}|>2\pi/l_{NC}$
can be interpreted as string like extended objects whose
length is $|C^{\mu\nu}k_{\nu}|$.
When $|k_{\mu}|<2\pi/l_{NC}$, on the other hand, this matrix
becomes close to diagonal whose matrix elements go like
\beq
\langle x_i|exp(ik\cdot\hat{x})|x_j \rangle \sim
exp(ik\cdot x_i) \langle x_i|x_j \rangle .
\label{diagonalpart}
\eeq
It again supports our interpretation that
$exp(ik\cdot\hat{x})$ correspond to the ordinary plane waves
when $|k_{\mu}|<2\pi/l_{NC}$. They are represented by the
matrices which are close to diagonal.
\par
\begin{figure}[h]
\begin{center}
\leavevmode
\epsfxsize=6cm
\epsfbox{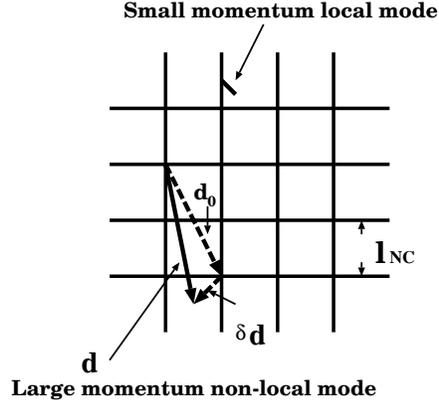}
\caption{
Noncommutative fields are represented as bi-local fields
 on the von Neumann lattice with
the lattice spacing $l_{NC}$.
 Small momentum modes ($|k|< 2 \pi/l_{NC}$)
represent ordinary (commutative) fields.
Large momentum
modes ($|k|> 2 \pi/l_{NC}$) represent bi-local `open strings'
which are highly nonlocal.
Only a fraction of the full momentum ($|k^c| < 2 \pi/l_{NC}$)
can be interpreted as the momentum associated with the center of mass
motion of an `open string' on the von Neumann lattice.
In the figure, $\delta d^{\mu}= C^{\mu \nu} k^c_{\nu}$.
}
\label{fig:lattice}
\end{center}
\end{figure}
We may expand matrices $\hat{\phi}$ in the twisted reduced model
by the following bi-local basis as follows:
\beq
\hat{\phi}=\sum_{i,j} \phi(x_i, x_j)|x_i \rangle \langle x_j| ,
\label{bi-local}
\eeq
where the Hermiticity
of $\hat{\phi}$ implies $\phi^*(x_j, x_i)=\phi(x_i, x_j)$.
The matrices $\hat{\phi}$ represent $\hat{a_{\mu}}$ or $\hat{\psi}$
in the super Yang-Mills case but the setting
here is more generally applied to
 an arbitrary  noncommutative field theory.
The bi-local basis spans the whole $n^2$ degrees of freedom of matrices.
\footnote{Bi-local fields have also appeared in $c=1$ string theory
\cite{Wadia}\cite{Sakita}.}
\par
Here we work out the translation rule between the
momentum eigenstate representation
$\hat{\phi}= \sum_k \tilde\phi(k) exp(ik\cdot\hat{x})$ and the bi-local
field representation of eq.(\ref{bi-local}):
\beqa
 \tilde\phi (k) &=& {1 \over n} Tr (exp(-ik\cdot\hat{x}) \hat{\phi})
= {1 \over n} \sum_{i,j}
\langle x_i| exp(-ik\cdot\hat{x}) |x_j \rangle \
\phi(x_j, x_i) \n
&=& {1 \over n} \sum_{x_c} \phi(x_c,d) exp(-ik \cdot x_c),\n
\phi(x_c,d)&=&\sum_{x_r} exp({i\over 2}B_{\mu\nu}x_r^{\mu}x_c^{\nu})
exp(-{(x_r-d)^2\over 4C})\phi (x_j,x_i),
\label{xcxr}
\eeqa
where $x_c=(x_i+x_j)/2$ and $x_r=x_i-x_j$. From eq.(\ref{xcxr}), we
observe that the slowly varying field with the momentum smaller than
$2\pi /l_{NC}$ consists of the almost diagonal components.
Hence close to diagonal components of the bi-local field
are identified with the ordinary slowly varying field $\phi(x_c)$.
On the other hand, rapidly oscillating fields are mapped to the
off-diagonal open string states.
A large momentum in the $\nu$-th direction
$|k_{\nu}| > 2\pi /l_{NC} $ corresponds to a large distance in the $\mu$-th
direction
$|d^{\mu}| = |C^{\mu \nu} k_{\nu}| > l_{NC}$.

We can decompose $d$ as $d = d_0 + \delta d$
where $d_0$ is a vector which connects two points on the von Neumann lattice
and $|\delta d^{\mu}| < l_{NC}$.
This decomposition is illustrated in Figure 2.
Then the summation over $x_r$ in (\ref{xcxr})
is dominated at $x_r = d_0$.
In this way
the large momentum degrees of freedom are more naturally interpreted
as extended open string-like fields.
They are denoted by `open strings' in this paper.
In this representation, we make contact with the quenched reduced models
\cite{RM} in the large momentum region.

Here we remark the important property
concerning the infra-red and ultra-violet cut-offs
of NCYM constructed with $n$ dimensional matrices.
Since the unit lattice size of
the von Neumann lattice is $l_{NC}$,
the total lattice size is $l_{NC} n^{1/\tilde{d}}$.
It implies that the maximum momentum is
$2\pi n^{1/\tilde{d}}/l_{NC}$ by using the relation
$\hat{p}_{\mu}=B_{\mu\nu}\hat{x}^{\nu}$.
It in turn implies that the minimum momentum
of the system is
$2\pi /n^{1/\tilde{d}}l_{NC}$ since we
have $n^2$ momentum modes.
The matrix model construction of NCYM
implies very natural infra-red and ultra-violet cut-offs
which disappear in the large $n$ limit.

\section{Estimations of the string scale}
\setcounter{equation}{0}
\hspace*{\parindent}
In this section, we estimate the string scale $\alpha '_{eff}$
in IIB matrix model with noncommutative backgrounds.
We have explained that the von Neumann lattice
naturally appears in the preceding section.
We argue that NCYM is superstring theory
on the von Neumann lattice.
We first give the arguments based on the tree level propagators.
We then explain that our claim is supported by T-duality arguments.
We give another evidence for it by investigating graviton exchange
processes in the next section.

As we have shown in the preceding section,
the momentum which can be associated with the center of mass
motion of an `open string'
is not full $k_{\mu}$ but rather $k^c_{\mu}=B_{\mu\nu}\delta d^{\nu}$.
There we have decomposed $d^{\mu}=C^{\mu\nu}k_{\nu}$ as $d = d_0 + \delta d$
where $d_0$ is a vector which connects two points on the von Neumann lattice
and $|\delta d^{\mu}| < l_{NC}$.
We can indeed represent $\tilde{\phi}(k)$ of eq.(\ref{xcxr})
as follows:
\beq
\tilde{\phi}(k^c,d)
={1 \over n} \sum_{x_c} \phi(x_c,d) exp(-ik^c \cdot x_c) .
\eeq
It is because
\beq
exp(-iB_{\mu\nu}d_0^{\nu}x_c^{\mu})=1 .
\eeq
We interpret $\tilde{\phi}(k^c,d)$ as the creation-annihilation
operator for the open string with momentum
$k^c$ and length $d$.

We consider the following tree level propagator:
\beq
\sum_{k_0}<\tilde{\phi} (-k)\tilde{\phi} (k)>
exp(ik_0\tau)
\sim exp(-m\tau)
\label{corel}
\eeq
where $k_{0}$ and $\vec{k}$ denote time-like and
spatial momenta respectively in this section.
The mass term is conventionally identified with the zero
spacial momentum limit of the correlator eq.(\ref{corel}).
In order to relate it to the mass of an `open string',
we consider a state with $k_1 <2\pi/l_{NC} < k_2,k_3$.
Such a state is extended in $(x^2,x^3)$ plane with the length
$l=C\sqrt{k_2^2+k_3^2}$.
As we have argued,
the momentum which can be associated with the center of mass
motion of an `open string'
is not full $\vec{k}$ but rather $\vec{k}^c$.
We find $m\sim Bl$ from the zero momentum limit
($\vec{k}^c \rightarrow 0$) of the correlator
eq.(\ref{corel}). From these considerations,
we propose to identify the mass of the state
with the length $l>>l_{NC}$ as
$Bl$.  We recall that an open string with the length $l$ has the mass of
$l/2\pi \alpha '$.
Therefore we find $2\pi \alpha '_{eff}=C$ with such an identification.

We remark that our estimate is consistent with
the string theory arguments. From eq.(\ref{openprop}),
we have indeed found that the $\alpha '_{eff}\sim c$
in section 2.
A hint for such an identification has been found in
a finite temperature investigation in \cite{Fischler}.
Our estimate is also consistent with the space-time
uncertainty principle of Yoneya\cite{yoneya}.

It is certainly true that we obtain
NCYM in the low energy limit
of string theory with $b_{\mu\nu}$ backgrounds.
In such a limit, we retain only those degrees of freedom
whose masses are smaller than the string scale.
However in IIB matrix model with noncommutative backgrounds,
we also find very high energy modes which may be interpreted
as long `open strings' due to the relation
$\hat{x}^{\mu}=C^{\mu\nu}\hat{p}_{\nu}$. As can be seen in eq.(\ref{openprop})
they are as massive as oscillator modes if their lengths exceed $l_{NC}$.
We therefore emphasize here that our formulation
is not a low energy limit of string theory.
The graviton exchanges are observed only because we have very
long `open strings' in the matrix model.
To put it differently, infrared singular
behaviors are observed in noncommutative field theory
only if we consider the
ultra-violet cut-off which is much larger than the
noncommutativity scale.
It is the reason why closed strings do not decouple
in such a formulation.

In our picture, we interpret the bi-local fields as
the zero modes of open strings.
We classify the zero modes as `momentum modes'
and `winding modes' as follows.
We recall the von Neumann lattice $|x_i>$
which is constructed
by the generators of the translation
operators $U_{\mu}=exp(-il_{NC}e_{\mu}\cdot \hat{p})$.
We can `compactify' the theory by
imposing the following conditions for fluctuations
\beq
\hat{\phi} =U_{\mu}\hat{\phi} U^{\dagger}_{\mu} .
\eeq
In this T-dual picture, the von Neumann lattice
can be identified with the lattice spanned by the
winding modes. We thus classify those modes as `winding modes'.
We have explained that $k^c_{\mu}$ can be interpreted
as momentum modes which can be associated with the center of mass
motion of `open strings' on the von Neumann lattice.
We thus classify them as `momentum modes'.

In string theory, the introduction of constant $b_{\mu\nu}$ background
is known to interpolate the Neumann and Dirichlet boundary
conditions. In the large and small $b_{\mu\nu}$ limit, we find
Dirichlet and Neumann boundary conditions respectively.
We find only `winding' and `momentum' modes in these limits.
If we expand IIB matrix model
around the commutative backgrounds, we only find `winding' modes.
In string theory we also find only `winding' modes
if we consider strings which connect D instantons.
The advantage of noncommutative backgrounds is that we find
the both `momentum' and `winding' modes.

Since we have found the both `momentum' and `winding' modes,
it is no surprise that the theory possesses T duality.
The remarkable property of NCYM
is the existence of Morita equivalent pairs
\cite{PS}\cite{SW}\cite{HI}.
We propose that two Morita equivalent theories can be
related by the exchange of the `momentum' and `winding'
modes.

We have argued that the `winding' modes of NCYM
span the von Neumann lattice whose lattice unit size is $l_{NC}$.
The total lattice size is $l_{NC} n^{1/\tilde{d}}$.
We may reinterpret it as the maximum momentum
$2\pi n^{1/\tilde{d}}/l_{NC}$ of the dual lattice.
The dual lattice possesses the unit lattice size of $l_{NC}/ n^{1/\tilde{d}}$.
We consider a twisted large $N$ reduced model on such a lattice.
In this way we find a pair of theories with the
compactification radii
$R=l_{NC}n^{1/\tilde{d}}/2\pi$ and $R'=l_{NC}/ n^{1/\tilde{d}}2\pi$.
They are related by the duality transformation
$R'=\alpha '_{eff}/R$ with $2\pi \alpha '_{eff}=C$.
Next we recall that the `momentum' modes
of NCYM are quantized in the unit of $l_{NC}/ n^{1/\tilde{d}}C$.
We can naturally reinterpret them as the `winding' modes of the dual lattice.
These winding modes can be
obtained by introducing the unit magnetic flux
in $U(n)$ gauge theory by imposing twisted boundary conditions.
In this sense NCYM and twisted reduced models are Morita equivalent
\cite{AMNS1}.

We remark that the T-duality transformation we have discussed
is expressed by the following open string metric transformation
in string theory\cite{SW}
\beq
G_{\mu\nu}\rightarrow \Theta^{\mu\rho} G_{\rho\sigma}(\Theta^T)^{\sigma\nu}
\label{tdual}
\eeq
where $\Theta^{\mu\nu}=C^{\mu\nu}/(2\pi R^2) \sim 1/n^{2/\tilde{d}}$.
Our interpretation is that
the two metrics which are related by the T-duality transformation
in eq.(\ref{tdual}) describe two tori we have just constructed.
We conclude that our estimate of the inverse string tension
$2\pi \alpha '_{eff}=C$ is
also supported by the T-duality arguments.
We argue that this is the exact result since it is
obtained from the exact T duality property of the theory.

\section{Graviton exchange processes}
\setcounter{equation}{0}

In this section, we study gravitational interactions in IIB matrix
model with noncommutative backgrounds.
To be specific, we consider photon-photon scattering
via exchange of a graviton.
This process can be studied by considering
block-block interactions. Namely we consider the backgrounds
of the following type:
\beqa
A_{\mu}&=&\left( \begin{array}{cc}
p_{\mu}+a_{\mu} & 0 \\
0 & p_{\mu}+a'_{\mu}
\end{array}
\right) ,
\eeqa
where $a_{\mu}$ and $a'_{\mu}$ denote the backgrounds
which represent two colliding photons.

The one-loop effective action
of IIB matrix model is
\beq
ReW  = \frac{1}{2}{\cal T}r \log(P_{\lambda}^2 \delta_{\mu\nu}-2iF_{\mu\nu})
      -\frac{1}{4}{\cal T}r
\log((P_{\lambda}^2+\frac{i}{2}F_{\mu\nu}\Gamma^{\mu\nu})
      (\frac{1+\Gamma_{11}}{2}))-{\cal T}r \log(P_{\lambda}^2).
\label{oneloopeffpot}
\eeq
Here $P_{\mu}$ and $F_{\mu\nu}$ are operators acting on the space of
matrices as
\beqa
        P_{\mu}X & = & [p_{\mu},X] ,\n
        F_{\mu\nu}X & = & \left[ f_{\mu\nu},X \right],
\label{adjointoperator}
\eeqa
where $f_{\mu \nu}=i[p_{\mu},p_{\nu}]$.
Now we expand the general expression of the one-loop effective action
(\ref{oneloopeffpot}) with respect to the inverse powers of
the relative distance between the two blocks.
We quote the general expression in what follows\cite{IKKT}:
\beqa
W    &=&-{\cal T}r
   \left(\frac{1}{P^2}F_{\mu\nu} \frac{1}{P^2}F_{\nu\lambda}
             \frac{1}{P^2}F_{\lambda\rho}\frac{1}{P^2}F_{\rho\mu} \right) \n
       &~& -2{\cal T}r
     \left(\frac{1}{P^2}F_{\mu\nu} \frac{1}{P^2}F_{\lambda\rho}
             \frac{1}{P^2}F_{\mu\rho}\frac{1}{P^2}F_{\lambda\nu} \right) \n
       &~&+\frac{1}{2}{\cal T}r
   \left(\frac{1}{P^2}F_{\mu\nu} \frac{1}{P^2}F_{\mu\nu}
             \frac{1}{P^2}F_{\lambda\rho}\frac{1}{P^2}F_{\lambda\rho} \right)
\n
      &~&+\frac{1}{4}{\cal T}r
  \left(\frac{1}{P^2}F_{\mu\nu} \frac{1}{P^2} F_{\lambda\rho}
             \frac{1}{P^2}F_{\mu\nu}\frac{1}
{P^2}F_{\lambda\rho} \right)
       +O((F_{\mu\nu})^5).
\label{Wexpansion2}
\eeqa
Since  $P_{\mu}$
and $F_{\mu\nu}$ act on the $(i,j)$ blocks independently, the one-loop
effective
action $W$ is expressed as the sum of contributions of the $(i,j)$ blocks
$W^{(i,j)}$. Therefore we may consider $W^{(i,j)}$
as the interaction between the $i$-th and $j$-th blocks.

Using (\ref{Wexpansion2}) we can easily evaluate $W^{(i,j)}$
to the leading order of $1/\sqrt{(d^{(i)}-d^{(j)})^2}$ as
\beqa
W^{(i,j)}&=&
{1\over r^8}
(-{\cal T}r^{(i,j)}
(F_{\mu\nu}F_{\nu\lambda}F_{\lambda\rho}F_{\rho\mu})
-2{\cal T}r^{(i,j)}
(F_{\mu\nu}F_{\lambda\rho}F_{\mu\rho}F_{\lambda\nu})\n
&&+{1\over 2}{\cal T}r^{(i,j)}
(F_{\mu\nu}F_{\mu\nu}F_{\lambda\rho}F_{\lambda\rho})
+{1\over 4}{\cal T}r^{(i,j)}
(F_{\mu\nu}F_{\lambda\rho}F_{\mu\nu}F_{\lambda\rho})
\n
&=&{3\over 2 r^8}
( -n_j \tilde{b}_8(f^{(i)})- n_i \tilde{b}_8(f^{(j)}) \n
&&
-8Tr(f_{\mu\nu}^{(i)}f_{\nu\sigma}^{(i)})
Tr(f_{\mu\rho}^{(j)}f_{\rho\sigma}^{(j)})
+Tr(f_{\mu\nu}^{(i)}f_{\mu\nu}^{(i)})
Tr(f_{\rho\sigma}^{(j)}f_{\rho\sigma}^{(j)})) ,
\label{blockin}
\eeqa
where
\beq
\tilde{b}_8(f)=
{2\over 3}(Tr(f_{\mu\nu}f_{\nu\lambda}f_{\lambda\rho}f_{\rho\mu})
+2Tr(f_{\mu\nu}f_{\lambda\rho}f_{\mu\rho}f_{\lambda\nu})
-{1\over 2}Tr(f_{\mu\nu}f_{\mu\nu}f_{\lambda\rho}f_{\lambda\rho})
-{1\over 4}Tr(f_{\mu\nu}f_{\lambda\rho}f_{\mu\nu}f_{\lambda\rho})).
\eeq

So the photon-photon scattering amplitude
which corresponds to nonplanar diagrams in noncommutative gauge
theory is
\beqa
&&{3\over 2 r^8}
(-8Tr(f_{\mu\nu}^{(i)}f_{\nu\sigma}^{(i)})
Tr(f_{\mu\rho}^{(j)}f_{\rho\sigma}^{(j)})
+Tr(f_{\mu\nu}^{(i)}f_{\mu\nu}^{(i)})
Tr(f_{\rho\sigma}^{(j)}f_{\rho\sigma}^{(j)})
)\n
&=&
{3\over 2}({1\over 2\pi })^{\tilde{d}}B^{\tilde{d}-8}
\int d^{\tilde{d}}x\int d^{\tilde{d}}y{1\over (x-y)^8}\n
&&\{-8tr(f_{\mu\nu}(x)f_{\nu\sigma}(x))
tr(f_{\mu\rho}(y)f_{\rho\sigma}(y))
+tr(f_{\mu\nu}(x)f^{\mu\nu}(x))tr(f_{\rho\sigma}(y)f^{\rho\sigma}(y)) \},
\label{gravint}
\eeqa
where we have used our mapping rule
eq. (\ref{momrule}).

We consider the scattering of two
incident photons with the wave functions
\beq
e(p)_{\mu}exp(ip\cdot x),
~~e(q)_{\mu}exp(iq\cdot x) ,
\eeq
where $p^2=q^2=0$ and $p\cdot e(p)=q\cdot e(q)=0$.
In this case
\beqa
f_{\mu\nu}(p)&=&
p_{\mu}e(p)_{\nu}-p_{\nu}e(p)_{\mu},\n
f_{\mu\rho}(p)f^{\rho\nu}(-q)
&=&
-p_{\mu}q\cdot e(p)e(-q)_{\nu}
-q_{\nu}p\cdot e(-q)e(p)_{\mu}\n
&&
+p_{\mu}q_{\nu}e(p)\cdot e(-q)
+e(p)_{\mu}e(-q)_{\nu}p\cdot q .
\eeqa
If we consider the forward scattering limit
$k=p-q \rightarrow 0$:
\beq
f_{\mu\rho}(p)f^{\rho\nu}(-q)\rightarrow
p_{\mu}p_{\nu},
\eeq
we find only graviton exchange
in this limit.
\beq
-{12}({1\over 2\pi })^{\tilde{d}}B^{\tilde{d}-8}
\int d^{\tilde{d}}x\int d^{\tilde{d}}y{1\over (x-y)^8}
tr(f_{\mu\rho}(x)f^{\rho\nu}(x))tr(f_{\mu\sigma}(y)f^{\sigma\nu}(y)) .
\label{gravint1}
\eeq

This expression reminds us the one photon exchange amplitude
between two conserved currents $j$ and $\tilde{j}$ in $QED_4$:
\beqa
&&\int {d^4k\over (2\pi )^4} j_{\mu}(-k){1\over
k^2+i\epsilon}\tilde{j}^{\mu}(k)\n &=&{i\over 4\pi}\int d^4x\int
d^4yj_{\mu}(x){1\over (x-y)^2-i\epsilon}\tilde{j}^{\mu}(y) .
\label{qedint}
\eeqa
We decompose currents into positive and negative frequency parts
\beqa
j_{\mu}(x)&=&j^+_{\mu}(x)+j^-_{\mu}(x),\n
j^+_{\mu}(x)&=&\int_0^{\infty} d\omega exp(-i\omega
x^0)j^+_{\mu}(\omega,\vec{x}),\n j^-_{\mu}(x)&=&\int_0^{\infty} d\omega
exp(i\omega x^0)j^-_{\mu}(\omega,\vec{x}) .
\eeqa
We rewrite eq.(\ref{qedint}) as follows
\beqa
&&{1\over 4}\int d^4x\int d\vec{y} j^-_{\mu}(x^0,\vec{x}){1\over
|\vec{x}-\vec{y}|}
 \tilde{j}^{+\mu}(x^0-|\vec{x}-\vec{y}|,\vec{y})\n
&&+{1\over 4}\int d^4y\int d\vec{x}
j^+_{\mu}(y^0-|\vec{x}-\vec{y}|,\vec{x}){1\over |\vec{x}-\vec{y}|}
 \tilde{j}^{-\mu}(y^0,\vec{y}) .
\eeqa
In this way we find retarded Lienard-Wiechert type interactions.
The point we would like to make here is that covariance implies
causality. Since ten dimensional covariance
is manifest in IIB matrix model, it naturally leads to
ten dimensional causality in Minkowski space-time.
On the other hand, ensuring ten dimensional causality
is highly nontrivial in $AdS$/CFT correspondence\cite{rey}.

We recall the relevant part of the  NCYM action as follows
\beqa
&& B^{{\tilde{d}\over 2}-4}({1\over 2\pi})^{\tilde{d}\over 2}
{1\over 4g^2}\int d^{\tilde{d}}x
 tr(
f_{\alpha\beta}f^{\alpha\beta})_{\star} .
\label{ncymact}
\eeqa
The energy momentum tensor can be
read off from it in the low energy limit as
\beq
B^{{\tilde{d}\over 2}-4}({1\over 2\pi})^{\tilde{d}\over 2}
{1\over g^2}
(f_{\mu\rho}f^{\rho\nu}-{1\over4}\delta_{\mu}^{\nu}
f_{\rho\sigma}f^{\sigma\rho}).
\eeq
So we can rewrite the gravitational interaction
eq.(\ref{gravint1}) as follows
\beq
-{12}g^4
\int d^{\tilde{d}}x\int d^{\tilde{d}}y{1\over (x-y)^8}
T_{\mu\nu}(x)T^{\mu\nu}(y) .
\label{gravint2}
\eeq

We recall the $\tilde{d}$ dimensional propagators
\beq
\int {d^{\tilde{d}}p \over (2\pi )^{\tilde{d}}}
{1\over p^2}exp(ip\cdot x)
={\Gamma ({d-2\over 2})\over 4\pi^{\tilde{d}\over 2}}{1\over x^{\tilde{d}-2}}.
\eeq
For $\tilde{d}=10$, we obtain
\beq
G_{10}(x)={3\over 2\pi^5}{1\over x^8} .
\eeq
The gravitational coupling $\kappa$ is found to be
\beq
\kappa^2 = 16\pi^5 g^4 .
\label{kappa}
\eeq
We also read off the $\tilde{d}$ dimensional Yang-Mills coupling from
eq.(\ref{ncymact}) as
\beq
g^2_{YM}=C^{(\tilde{d}-8)/2}g^2(2\pi)^{\tilde{d}/2} .
\label{barecoupl}
\eeq

Here we quote string theory predictions:
\beq
\kappa^2={1\over 2}(2\pi )^7g_s^2\alpha'^4,
~~g^2_{YM}=(2\pi )^{\tilde{d}-3} g_s
\alpha'^{(\tilde{d}-4)/ 2}.
\label{stringpre}
\eeq
Eq.(\ref{stringpre}) agrees with eq.(\ref{kappa}) and
eq.(\ref{barecoupl}) with our identification $2\pi \alpha'_{eff}=C$.
We also find that the IIB matrix model coupling
can be expressed by $\alpha '_{eff}$ and $g_s$ as
\beq
g^2=(2\pi )g_s{\alpha '_{eff}}^2 .
\eeq
which is consistent with our previous estimate through D-strings\cite{IKKT}.
What these investigations indicate is that NCYM
is superstring theory with the above identified string scale
and string coupling. We have argued that it is superstring theory
on the von Neumann lattice. Since the lattice spacing $l_{NC}$
is not visible in the low energy limit, it may be expected that it
behaves like ordinary superstring theory in the low energy limit.


We find that the gravitational interaction eq.(\ref{gravint2})
exhibits the identical power law behavior $1/(x-y)^8$ irrespective
of the dimensionality of the backgrounds $\tilde{d}$.
It appears as if these background represent D-branes
in flat ten dimensional space-time.
However we argue that such an interpretation
is premature since we have only considered the one loop effect here.
We argue that a more reliable picture is obtained through supergravity
approach which allows us to investigate nonperturbative effects.


As it is explained in section 2, we find Newton's force law
with these backgrounds.
This is due to the existence of a massless bound state
a la Randall and Sundrum.
Such an effect is not visible in the perturbative
calculations in this section.
Therefore the supurgravity description
of $NCYM_4$ suggests a nonperturbative compactification
mechanism in IIB superstring and matrix model.

In the matrix model construction,
the longitudinal size of the system
is bounded by $l_{NC}n^{1/4}$.
It also implies that the transversal size
$r$ is bounded by $l_{NC}n^{1/4}$ since
it is identified with the maximum energy scale of the system
(multiplied by $l_{NC}^2$).
In eq.(\ref{adsrs}),
the dilaton expectation value is $O(1)$ at the noncommutativity scale
$r^2\sim 1$.
We then find $g_{\infty}$ is $O(1/n)$ since the dilaton decays like
$1/r^4$ beyond the noncommutativity scale.
We have fixed $\lambda=g_{\infty}\alpha'^2b^2$ to be $O(1)$ which implies
that $\alpha'b \sim \sqrt{n}$. We conclude that $l_{NC}\sim l_s/n^{1/4}$
and $r$ never exceeds $O(l_s)$ where $l_s$ is the string scale.
Therefore there is simply no region with $r>l_s$ in the
matrix model.
We have taken the noncommutativity scale
to be $O(1)$ and the cuff-off scale of $r$ becomes
$O(n^{1/4})$. The cut-off can be removed in the large
$n$ limit of the matrix model construction.
In this way we can realize the entire space-time
which is described by eq.(\ref{adsrs}).

\section{Conclusions and Discussions}
In this paper we have argued that we can obtain Newton's
force law with $NCYM_4$.
Since it contains four dimensional gauge theory and gravitation,
it is a candidate of the unified theory.
It can be regarded as a compactification of ten dimensional IIB superstring
straight down to four dimensions.
It is naturally obtained in IIB matrix model
with noncommutative backgrounds.
Therefore it provides a nonperturbative
compactification mechanism of matrix models.

We have identified the bi-local fields as the zero
modes of open strings.
They can be interpreted as the `momemtum' and `winding' modes
on the von Neumann lattice.
Our identification of the effective string scale with the
noncommutativity scale is consistent with the exact T-duality
which interchanges the `momentum' and `winding' modes.
Although we have identified the zero
modes of open strings, we have not constructed oscillator modes.
We expect to find them in higher order diagrams.
Let us consider a propagator (ribbon diagram). We associate it with a bi-local
field since the double lines of the ribbon are mapped to
two distinct space-time points.
We need to draw many loops inside
the ribbon at higher orders in perturbation theory.
We can assign a space-time
point to each loop. Our conjecture is that such an object can be
interpreted as the propagator of oscillation modes.
These arguments are illustrated in Figures 3 and 4.

It is important to recall here that we identify the string coupling
with the topological expansion parameter of the Feynman diagrams
of NCYM. It is not equal to the NCYM coupling although the
both are related in the low energy limit as suggested by supergravity
solutions. The tree level string propagator is obtained by summing all
planar diagrams.
Our proposal that NCYM with maximal SUSY
may be interpreted as superstring theory on the von Neumann lattice
should be understood in this context.

\begin{figure}[h]
\begin{center}
\leavevmode
\epsfxsize=6cm
\epsfbox{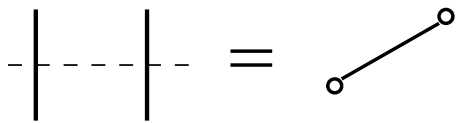}
\caption{Propagators of bi-local fields.
The two end-points are mapped onto space-time coordinates.
}
\label{fig:tree}
\end{center}
\end{figure}
\begin{figure}[h]
\begin{center}
\leavevmode
\epsfxsize=6cm
\epsfbox{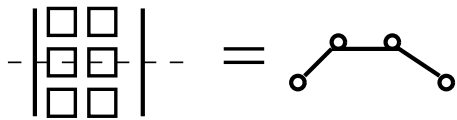}
\caption{Higher order corrections to propagators of bi-local fields
may render the bi-local fields stringy.
}
\label{fig:string}
\end{center}
\end{figure}

As we have pointed out, NCYM is obtained with a particular
classical background in IIB matrix model.
IIB matrix model is postulated as a nonperturbative
formulation of superstring theory.
In our proposal, the matrices
$A_{\mu}$ are to be interpreted as space-time coordinates.
If so, $\tilde{d}$ dimensional distributions of eigenvalues of
matrices represent $\tilde{d}$ dimensional space-time.
It is then expected that we find $\tilde{d}$
dimensional gauge theory and gravitation with
such a background.
In this paper we have argued that it is indeed the case
with maximally supersymmetric backgrounds. From the findings in this
paper, we draw the conclusion that NCYM provides a strong support for
our basic premises of our IIB matrix model conjecture.


\begin{center} \begin{large}
Acknowledgments
\end{large} \end{center}
This work is supported in part by the Grant-in-Aid for Scientific
Research from the Ministry of Education, Science and Culture of Japan.

\setcounter{equation}{0}

\end{document}